\documentclass{article}
\usepackage[T1]{fontenc} 
\usepackage[utf8]{inputenc} 
\usepackage{ismir,amsmath,cite,url}
\usepackage{graphicx}

\usepackage{color}

\usepackage[firstpage]{draftwatermark}
\definecolor{lightgray}{rgb}{0.9,0.9,0.9}
\definecolor{darkgray}{rgb}{0.4,0.4,0.4}
\SetWatermarkFontSize{12pt}
\SetWatermarkScale{1.1}
\SetWatermarkAngle{90}
\SetWatermarkHorCenter{202mm}
\SetWatermarkVerCenter{170mm}
\SetWatermarkColor{darkgray}
\SetWatermarkText{Late-Breaking / Demo Session Extended Abstract, ISMIR 2024 Conference}

\def\lbd{}	

\title{Demo of Zero-Shot Guitar Amplifier Modelling: Enhancing Modeling with Hyper Neural Networks}

\multauthor
{Yu-Hua Chen$^{1,3}$ \hspace{1cm} Yuan-Chiao Cheng$^3$  \hspace{1cm} Yen-Tung Yeh$^{2,3}$} { \bfseries{Jui-Te Wu$^3$ \hspace{1cm} Yu-Hsiang Ho$^3$ \hspace{1cm} Jyh-Shing Roger Jang$^1$ \hspace{1cm} Yi-Hsuan Yang$^2$}\\
$^1$ Graduate Institute of Networking and Multimedia, National Taiwan University, Taiwan\\
$^2$  Department of Electrical Engineering, National Taiwan University, Taiwan\\
$^3$ Positive Grid, Taiwan
\\{\tt\small f08946011@ntu.edu.tw, yhyangtw@ntu.edu.tw}
}

\def\authorname{Yu-Hua Chen, Yuan-Chiao Cheng, Yen-Tung Yeh, Jui-Te Wu, Yu-Hsiang Ho, Jyh-Shing Roger Jang, and Yi-Hsuan Yang}

\begin{document}

\maketitle
\begin{abstract}
Electric guitar tone modeling typically focuses on the non-linear transformation from clean to amplifier-rendered audio. Traditional methods rely on one-to-one mappings, incorporating device parameters into neural models to replicate specific amplifiers. However, these methods are limited by the need for specific training data. 
In this paper, we adapt a model based on the previous work, which leverages a tone embedding encoder and a feature wise linear modulation (FiLM) condition method. In this work, we altered conditioning method using a hypernetwork-based gated convolutional network (GCN) to generate audio that blends clean input with the tone characteristics of reference audio.
By extending the training data to cover a wider variety of amplifier tones, our model is able to capture a broader range of tones. Additionally, we developed a real-time plugin to demonstrate the system's practical application, allowing users to experience its performance interactively. Our results indicate that the proposed system achieves superior tone modeling versatility compared to traditional methods.

\end{abstract}
\section{Introduction}\label{sec:introduction}
 \begin{figure}
 \centerline{{
 \includegraphics[alt={ISMIR 2024 template test image},width=1.0\columnwidth]{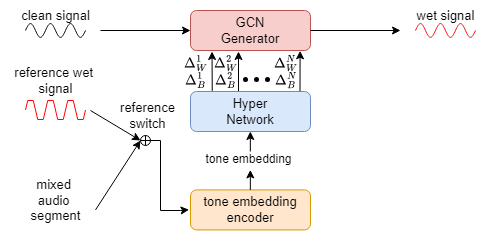}}}
 \caption{Diagram of our system Workflow. A reference audio can either be a wet signal or a guitar recording extracted from a mixed audio segment found on public platforms (e.g., YouTube).}
 \label{fig:example}
\end{figure}

Electric guitar tone modeling focuses on the non-linear transformation between clean and amplifier-rendered audio. Several networks have been proposed for emulating guitar amplifiers, either in controlled settings with adjustable parameters or in snapshot tone capture scenarios. Most previous works \cite{damskagg2019deep, wright2019real} have concentrated on modeling one-to-one mappings, where device parameters are incorporated into a neural model to fully emulate a specific amplifier. Other approaches \cite{wright2023adversarial, chen24dafx}, which use generative adversarial networks (GANs) to generate tones from unpaired data (i.e., content and tone are unpaired), have extended amplifier modeling to more flexible training scenarios.
A related area of interest is singing voice conversion (SVC)\cite{deng2020pitchnet,takahashi2021hierarchical,liu2021fastsvc}, where several applications have been proposed. However, both open-source and commercial products in this domain still require training data specific to the target speaker or singer, resulting in limited flexibility. This challenge parallels the limitations in guitar tone modeling, where the need for specific training data constrains adaptability.

Building on the concept of speaker embedding in SVC, recent studies have introduced a promising approach that captures and represents amplifier tones from reference audio using an embedding vector. This allows for the rendering of clean audio with the tone characteristics of the reference audio. This raises an important question: if a generator is trained on a broader variety of tones, encompassing almost all types of commercial amplifiers, could it achieve more accurate zero-shot tone modeling? We argue that a more flexible representation of "tone" can be used as a condition for a model to replicate the amplifier tone of a referenced audio in real-world usage.

The task can be defined as follows: given a clean audio sample and a reference audio, the tone embedding is extracted by a tone embedding encoder, allowing us to generate an output that combines the content of the clean audio with the tone of the reference audio. The conditioning mechanism is crucial in this process. Previous studies have proposed using TCNs and hypernetworks for multiple models across several devices. In \cite{chen2024towards}, it was found that gated convolutional networks (GCN) outperforms look-up table conditioning approaches. Furthermore, \cite{yeh2024hyperrecurrentneuralnetwork} compared different hypernetworks for modeling various analog devices and demonstrated that hypernetworks exhibit superior performance and computational efficiency compared to FiLM\cite{perez2018film}.

In this study, we propose a system based on our previous work \cite{chen2024towards} and extend the scope of the training data by increasing both the duration and the number of effects in the dataset. The diagram for our system is shown in figure \ref{fig:example}.
By adapting the model to a hypernetwork-based GCN, we have also implemented a C++ version of the plugin with a NAM interface. Additionally, our system allows the reference audio to be sourced from a mixed audio segment from public platforms, such as YouTube. Users can select any arbitrary segment, which is then processed through our internal source separation model. The resulting guitar-only segment is used as the reference audio for our model.

\section{Implementation Details}

In \cite{chen2024towards}, we employed a GCN \cite{rethage2018wavenet} as their backbone model, incorporating feature-wise linear modulation (FiLM) \cite{perez2018film} for conditioning. We also adopt this same backbone architecture as the foundation of our generator model. In the domain of conditional audio synthesis, \cite{richard2021neural} utilized a hypernetwork \cite{ha2016hypernetworks} to integrate conditioning information, generating weights for the layers in a convolutional model for mono-to-binaural synthesis. When applying hypernetworks to recurrent neural networks (RNNs), \cite{yeh2024hyperrecurrentneuralnetwork} found that dynamicHyper-GRU achieved superior performance across several metrics compared to FiLM and concatenation-based conditioning mechanisms, particularly in the context of parameter conditioning on the Boss OD-3 pedal.

To further explore which embedding-conditioned model is most suitable for zero-shot tone modeling, we incorporate a hypernetwork into our model to compare its performance against the combination of GCN and FiLM. Our hypernetwork is composed of 20 hyper blocks, where each hyper block contains 
3 hyper layers. Each hyper layer takes the tone embedding as a conditional input to generate deltas for the weights and biases of the convolutional layers in each GCN.

The workflow of our sysem is: given a clean audio signal \( \mathbf{x} \) and a referenced audio signal \( \mathbf{x}_{\text{ref}} \), 
we first extract the tone embedding \( \phi \) using a tone embedding encoder model \(\mathcal{E} \), which takes the referenced audio as input.

Next, the Hypernetwork \( \mathcal{H}_l \) takes this extracted tone embedding to generate the delta of weights \( \Delta\mathbf{W}_l \) and delta of biases \( \Delta\mathbf{b}_l \) for each original weight  \(\mathbf{W}_l \) and biases \( \mathbf{b}_l \)  of convolutional layer \( l \) of the generator.

Finally, the clean audio \( \mathbf{x} \) undergoes convolution operations through every layer of the generator \( G\) to produce the rendered result \( \mathbf{y} \).

\begin{gather*}
\mathcal{\phi} = \mathcal{E}(\mathbf{x}_{\text{ref}}) \\[1em]
\Delta\mathbf{W}_l, \Delta\mathbf{b}_l = \mathcal{H}_l(\phi) \\[1em]
\mathbf{W}_l^{\text{new}} = \mathbf{W}_l \cdot (1 + \Delta \mathbf{W}_l), \quad 
\mathbf{b}_l^{\text{new}} = \mathbf{b}_l \cdot (1 + \Delta \mathbf{b}_l) \\[1em]
\mathbf{y} = G(\mathbf{x}; \{\mathbf{W}_l^{\text{new}}, \mathbf{b}_l^{\text{new}} \}_{l \in L})
\end{gather*}

Throughout the training process, the tone embedding model  \( \mathcal{E}\) remains fixed.

\subsection{Dataset}\label{sec:page_size}

Since the model in \cite{chen2024towards} was only trained on nine amplifiers (three each for high-gain, low-gain, and crunch types), it is challenging to assert that the generator can accurately model a wide range of amplifier tones. To address this limitation, we collaborated with a guitar amp and effects modeling company Positive Grid to significantly expand the diversity and quantity of amplifier types in our training data. Our dataset now includes nearly all possible combinations of head and cabinet configurations available in the BIAS FX2 plugin, resulting in 80000 distinct tones and covering a total duration of 5300 hours. All audio samples are formatted at 44.1 kHz to ensure compatibility with professional recording environments.




\section{Real Time Plugin}
To validate the effectiveness of zero-shot tone modeling, we implemented a real-time plug-in that allows users to experience the dynamic feedback and tactile response of the model, in addition to simply listening to the generated results. We referenced the Neural Amp Modeler (NAM), an open-source product also based on WaveNet, and modified its core DSP library \footnote{https://github.com/Tr3m/namcore-old/tree/main} to align with our model's requirements. The plug-in was developed using the JUCE framework\footnote{https://juce.com}, leveraging on the open-source resources from NAM JUCE \footnote{https://github.com/Tr3m/nam-juce} for demonstration purposes. Similar to the widely adopted NAM product, after empirical testing, our model requires only a comparable amount of computational power, enabling it to run efficiently on most computer systems without GPU resource.

\section{Conclusion}

This study introduces a novel approach to guitar tone modeling using a hypernetwork-based GCN model for zero-shot tone modeling on guitar amplifiers. Our system offers flexible and accurate tone rendering across various amplifier types. Expanded training data and a real-time plugin enhance its practical utility. The proposed system provides a robust solution for diverse tone modeling. Future work may focus on further optimizations and extending the framework to other effects.

\bibliography{ISMIRtemplate}

%
%
%
%
%

\end{document}